\documentclass{PoS}

\usepackage{epsfig,graphicx}
\usepackage{amsmath,amsfonts,amssymb}

\title{Phenomenological viability of string inspired multi-Higgs doublet
models}

\ShortTitle{Phenomenological viability of string inspired multi-Higgs doublet
models}

\author{\speaker{Ana Teixeira}\thanks{Based on work done in
        collaboration with C. Mu\~noz and N. Escudero. Work supported
        by ``Funda\c c\~ao para a Ci\^encia e Tecnologia'', under
        grant SFRH/BPD/11509/2002.}\\
        Departamento de F\'{\i }sica Te\'{o}rica C-XI and Instituto
de F\'{\i }sica Te\'{o}rica C-XVI, \\
        Universidad Aut\'{o}noma de Madrid,\\
        Cantoblanco, E-28049 Madrid, Spain\\
        E-mail: \email{teixeira@delta.ft.uam.es}}

\abstract{
We analyse the phenomenological viability of 
heterotic $Z_3$ orbifold models 
with two Wilson lines, which naturally predict three families
of matter and Higgs fields.
We study the orbifold parameter space, and discuss the compatibility of the
predicted Yukawa couplings with current experimental data, thus
evaluating the viability of the orbifold configurations.
We address the implications of tree-level flavour changing neutral processes
in constraining the Higgs sector of the model, finding that viable scenarios
can be obtained for a fairly light Higgs spectrum.
}

\FullConference{International Europhysics Conference on High Energy Physics\\
                 July 21st - 27th 2005\\
                 Lisboa, Portugal}

\begin{document}

\section{Introduction}
Even though the standard model (SM)
has proved to offer a very successful description of strong and
electroweak interactions, it fails in providing an explanation to
issues such as the gauge group, the number of families, the
dynamics of flavour and the mechanism of mass generation, among others.
On the topic of the structure of fermion masses and mixings, string
theory provides some hints towards a natural explanation. The particular
case of the abelian $Z_3$ orbifold compactification of the heterotic
superstring with two Wilson lines 
is especially appealing, since it predicts the 
standard model (SM) fermion content and correct gauge group, and offers
a scenario of explicitly computable and renormalizable Yukawa
couplings~\cite{Abel:2002ih}. 
The latter receive exponential suppression factors
that dependend on the distance between the fixed orbifold points to
which the fields are attached. In this case, family replication occurs
not only in the fermions, but also in the Higgs sector, and one has
three families of Higgs doublets. This is a direct consequence of the
orbifold compactification: since we have 27 fixed points, the twisted
matter comes in 9 sets with 3 equivalent sectors in each.
Having six Higgs doublets coupling to matter with distinct Yukawa
couplings potentially induces dangerous flavour changing neutral
currents (FCNCs) at the tree level. Therefore, the viabilty of this
class of models depends not only in accommodating the current data on
quark masses and mixings, but also on being in agreement with the
experimental data on neutral meson mass differences (e.g. $\Delta
m_K$)~\cite{Escudero:2005:2}.

\section{Yukawa couplings in $Z_3$ orbifold models with two Wilson lines}
After Fayet-Iliopoulos (FI) breaking, the SM matter survives as the
 massless mode of the original Yukawa coulings. The mass matrices of the
low-energy effective theory are now modified:
\begin{equation}\label{quark:mass}
\mathcal{M}^u= 
g \,N \,a^{u^c} \,A^u \,B^{u^c}\,, \quad \quad
\mathcal{M}^d= 
g \,N \,\varepsilon_1 \,a^{d^c}\, A^d\, B^{d^c}\,,
\end{equation}
where $g$ is the gauge coupling, $N$ is related to the volume of the
$Z_3$ lattice unit cell, $\varepsilon_i$ are exponential suppression
factors related to the shape and size of the orbifold ($\varepsilon_i
\propto \exp^{-\frac{2 \pi}{3} T_i}$, with $T_i$ the lattice diagonal moduli), 
and $a^f$ are functions of the VEVs of the SM singlet scalar fields
that cancel the FI $D$-term. The flavour content is encoded in 
\begin{equation}\label{quark:aFI}
A^u B^{u^c}= \left(
\begin{array}{ccc}
w_2\,\varepsilon_5 \, \beta^{u^c}& w_6 \,\varepsilon_5 & w_4 \, \alpha^{u^c} \\
w_6 \,\varepsilon_5^2 \, \beta^{u^c}& w_4 & w_2\, \alpha^{u^c} \, \\
w_4 \,\varepsilon_5^2 \, \beta^{u^c}& w_2\,\varepsilon_5 
& w_6\, \alpha^{u^c}/\varepsilon_5
\end{array}\right)\, , \,\,
A^d B^{d^c}= \left(
\begin{array}{ccc}
w_1\,\varepsilon_5 \, \beta^{d^c}& w_5 \,\varepsilon_5 & w_3 \, \alpha^{d^c} \\
w_5 \,\varepsilon_5^2 \, \beta^{d^c}& w_3 & w_1\, \alpha^{d^c} \, \\
w_3 \,\varepsilon_5^2 \, \beta^{d^c}& w_1\,\varepsilon_5 
& w_5\, \alpha^{d^c}/\varepsilon_5
\end{array}\right)\,.
\end{equation}
In the above, $\alpha^f$ and $\beta^f$ are also functions of the
singlet field VEVs, and of $\varepsilon_5$. $w_i$ are the VEVs of the
neutral components of the six Higgs doublets.
By comparing the eigenvalues of the above mass matrices (which can be
analytically obtained by expanding to first order in $\varepsilon_5$,
one can derive relations for the Higgs VEVs in terms of the down quark
masses:
\begin{equation}
\{w_1,w_3,w_5\} \propto \left\{
\frac{1}{\varepsilon_5 \beta^{d^c}} \left(m_d + \varepsilon_5^5
\frac{m_b^2}{m_s}\right), m_s, \frac{m_b \varepsilon_5}{\alpha^{d^c}}
\right\}\,,
\end{equation}
and likewise for the up-type VEVs, in relation to the with up-quark masses.
For given values of the quark masses, one fixes the ratio of the
several Higgs VEVs, which in turn allows to reconstruct the full quark
mass matrices, and obtain the quark spectra and mixings. Satisfying
the current experimental bounds on the latter, imposes severe
correlations on the orbifold parameters, $\{\varepsilon_5, \alpha^f,
\beta^f\}$. 
Taking 4 distinct sets of input quark masses, $\{ m_u, m_d, m_c, m_s,
m_t, m_b \}$ (in GeV) as 
\begin{equation}
\begin{array}{cc}
\begin{array}{lcccccc}
\mathrm{\underline{A}}:\ \ & 0.004  & 0.008 & 1.35 & 0.13  & 180 & 4.4 \\
\mathrm{\underline{C}}: \ \ & 0.0035 & 0.004 & 1.15 & 0.08  & 176 & 4.1 
\end{array}
\hspace*{3mm}&\hspace*{3mm}
\begin{array}{lcccccc}
\mathrm{\underline{B}}: \ \ & 0.0035 & 0.008 & 1.25 & 0.1   & 178 & 4.5 \\
\mathrm{\underline{D}}: \ \ & 0.004  & 0.006 & 1.2  & 0.105 & 178 & 4.25 
\end{array}
\end{array}
\end{equation}
and requiring that the points are in agreement with viable quark
masses and current $|V_{\mathrm{CKM}}|$ bounds \cite{pdg2004}, 
one obtains the following correlations on the orbifold parameters:
\begin{figure}[htb]
  \begin{center} \vspace*{-8mm}\hspace*{-10mm}
    \begin{tabular}{cc}
\psfig{file=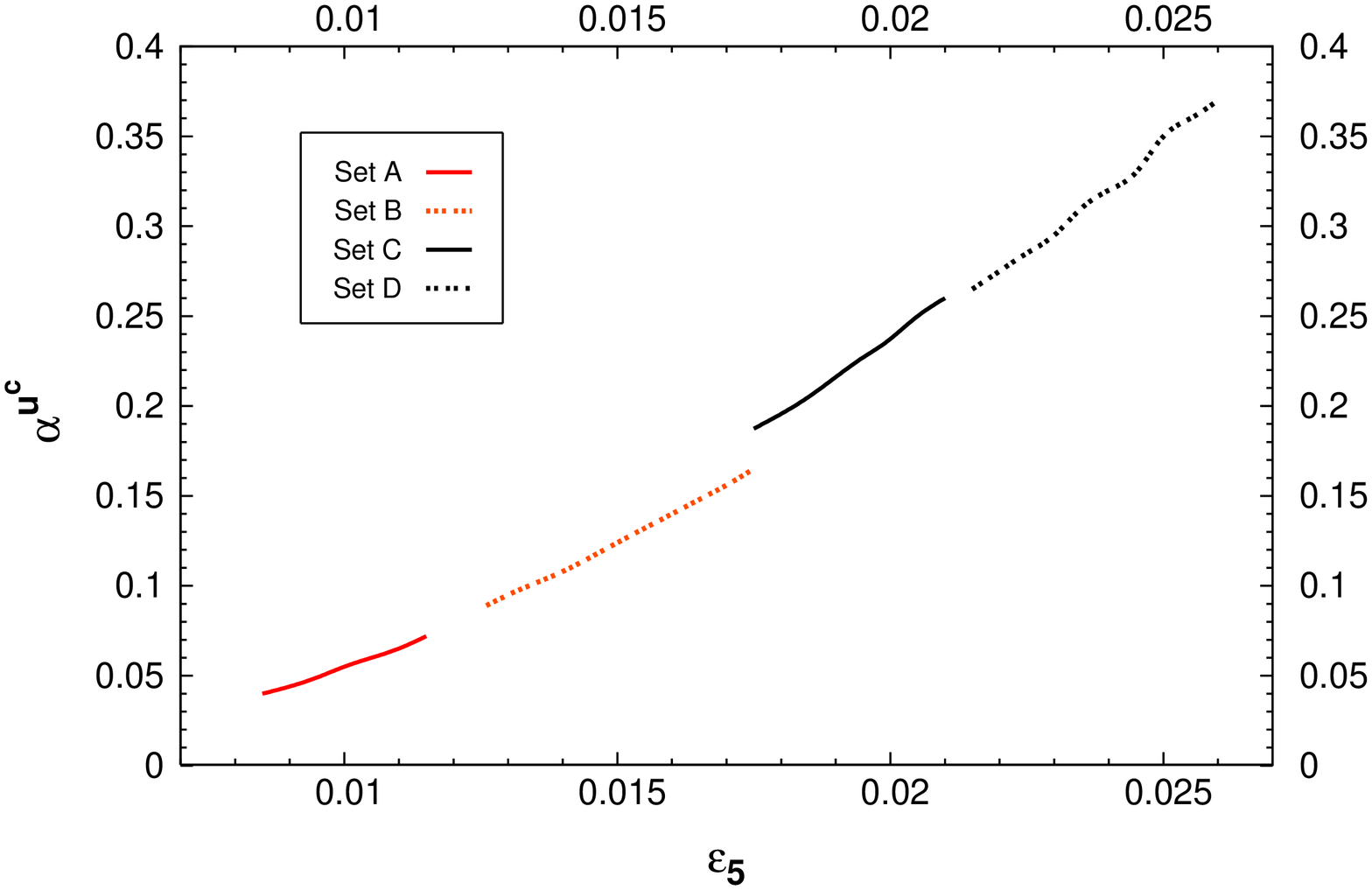,width=50mm,angle=270,clip=} &
\psfig{file=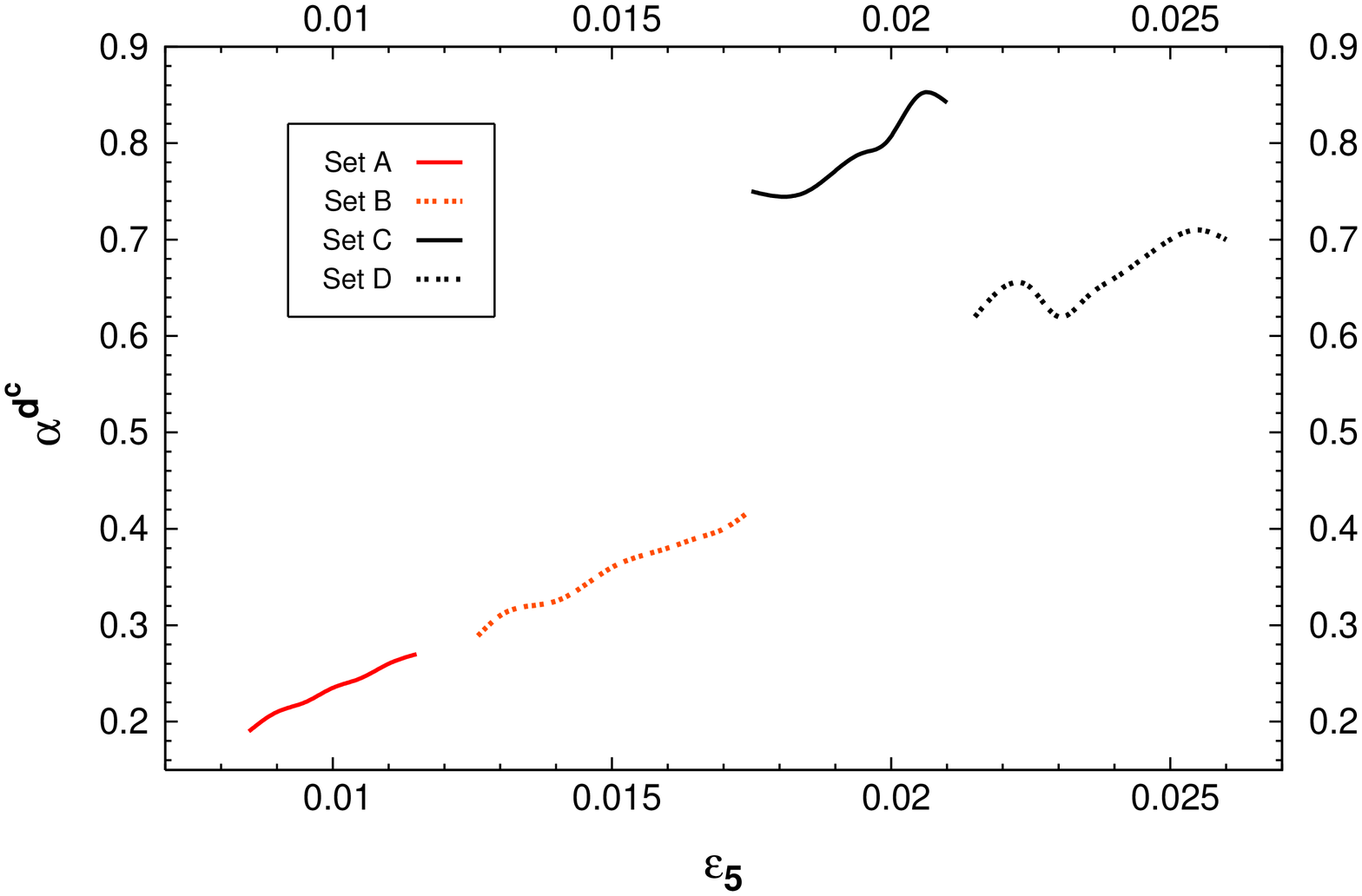,width=50mm,angle=270,clip=} 
    \end{tabular}
    \caption{Correlation between the orbifold parameters
($\alpha^{u^c}$, $\varepsilon_5$) and ($\alpha^{d^c}$,
    $\varepsilon_5$) for Sets A-D.} 
    \label{fig:orbifold:alpha:ud:e5}
  \end{center}\vspace*{-3mm}
\end{figure}
In addition to the hierarchy constraint imposed by the observed pattern
of quark masses, the VEVs must further comply with other constraints
as those arising from electroweak symmetry breaking, as $\sum w_i^2= 
2 \,M_Z^2/(g^2+g'^2)
\approx (174\text{ GeV})^2$, and from defining $\tan \beta$ as 
$\tan
\beta={\sqrt{w_2^2+w_4^2+w_6^2}}\,/\,{\sqrt{w_1^2+w_3^2+w_5^2}}$. 
Bringing together the several constraints, one can derive relations
that allow to extract the remaining orbifold parameters, $a^{u^c}$,   
$\varepsilon_1 a^{d^c}$, and thus $\varepsilon_1$ and the moduli
$T_{1,5}$.
In particular, requiring that $\varepsilon_1$ is a perturbative
parameter will disfavour regimes of large $\tan \beta$, as is clear
from the Figure below.
\begin{figure}[htb]
  \begin{center} \vspace*{-8mm}\hspace*{-10mm}
    \begin{tabular}{cc}
\psfig{file=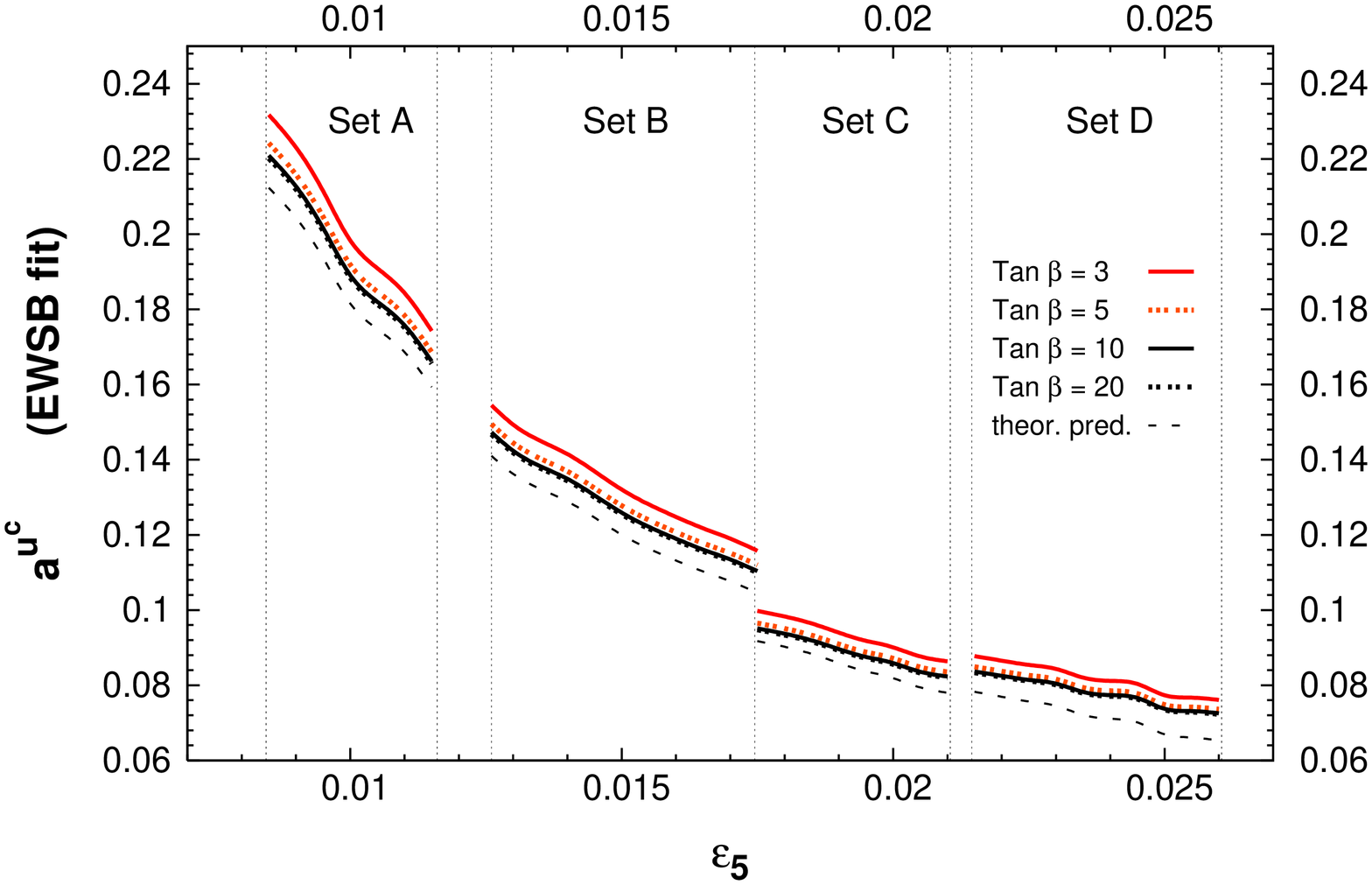,width=50mm,angle=270,clip=} &
\psfig{file=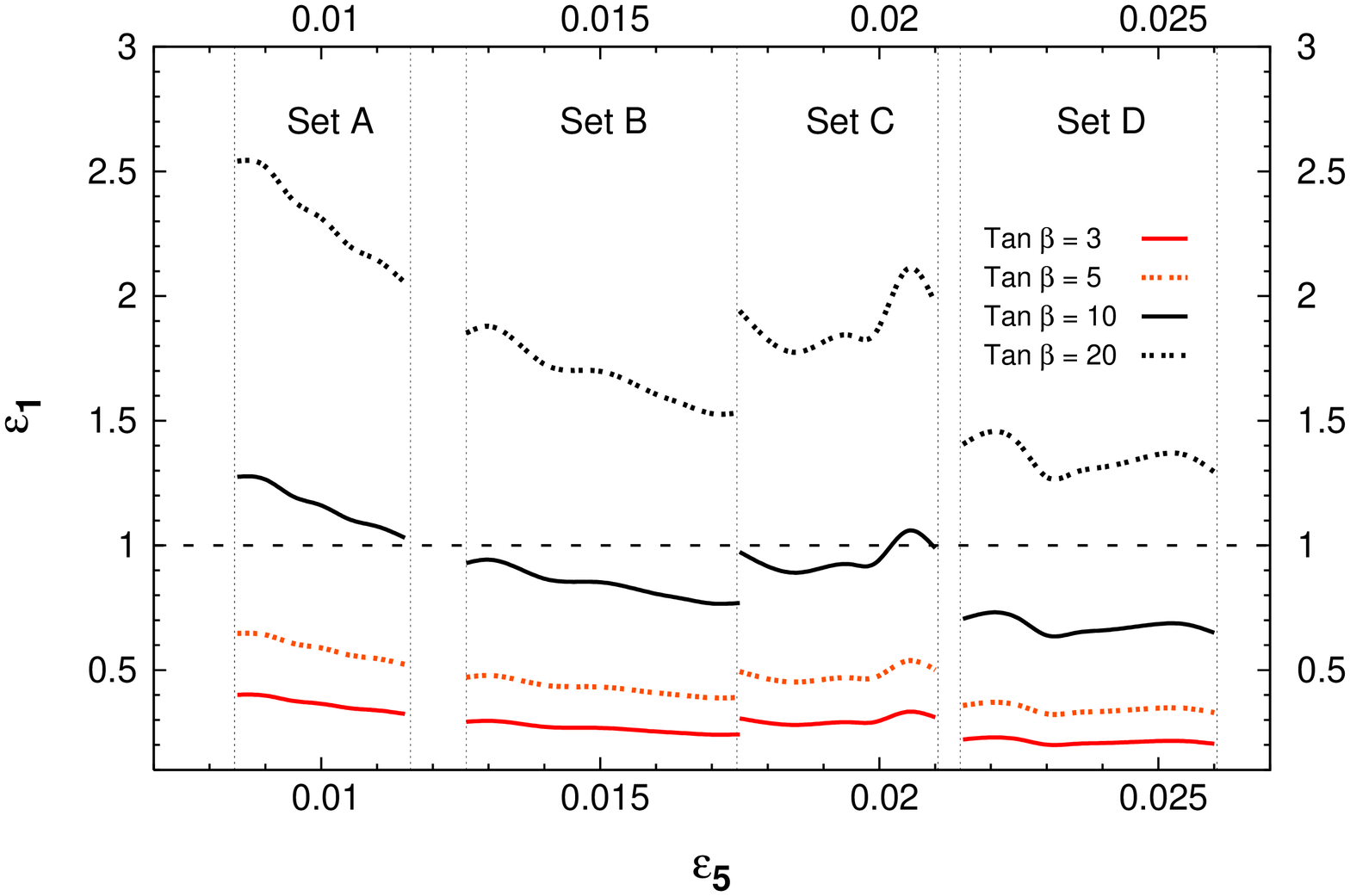,width=50mm,angle=270,clip=} 
    \end{tabular}
    \caption{EW symmetry breaking predictions and theoretical
    estimates of $a^{u^c}$ (left), and EW breaking estimate of
    $\varepsilon_1$ (right) for Sets A-D and $\tan \beta=3,5,10,20$.} 
    \label{}
  \end{center}\vspace*{-3mm}
\end{figure}
\section{An extended Higgs sector and FCNCs at the tree level}
\begin{figure}[htb]
  \begin{center} \vspace*{-8mm}\hspace*{-10mm}
    \begin{tabular}{cc}
\psfig{file=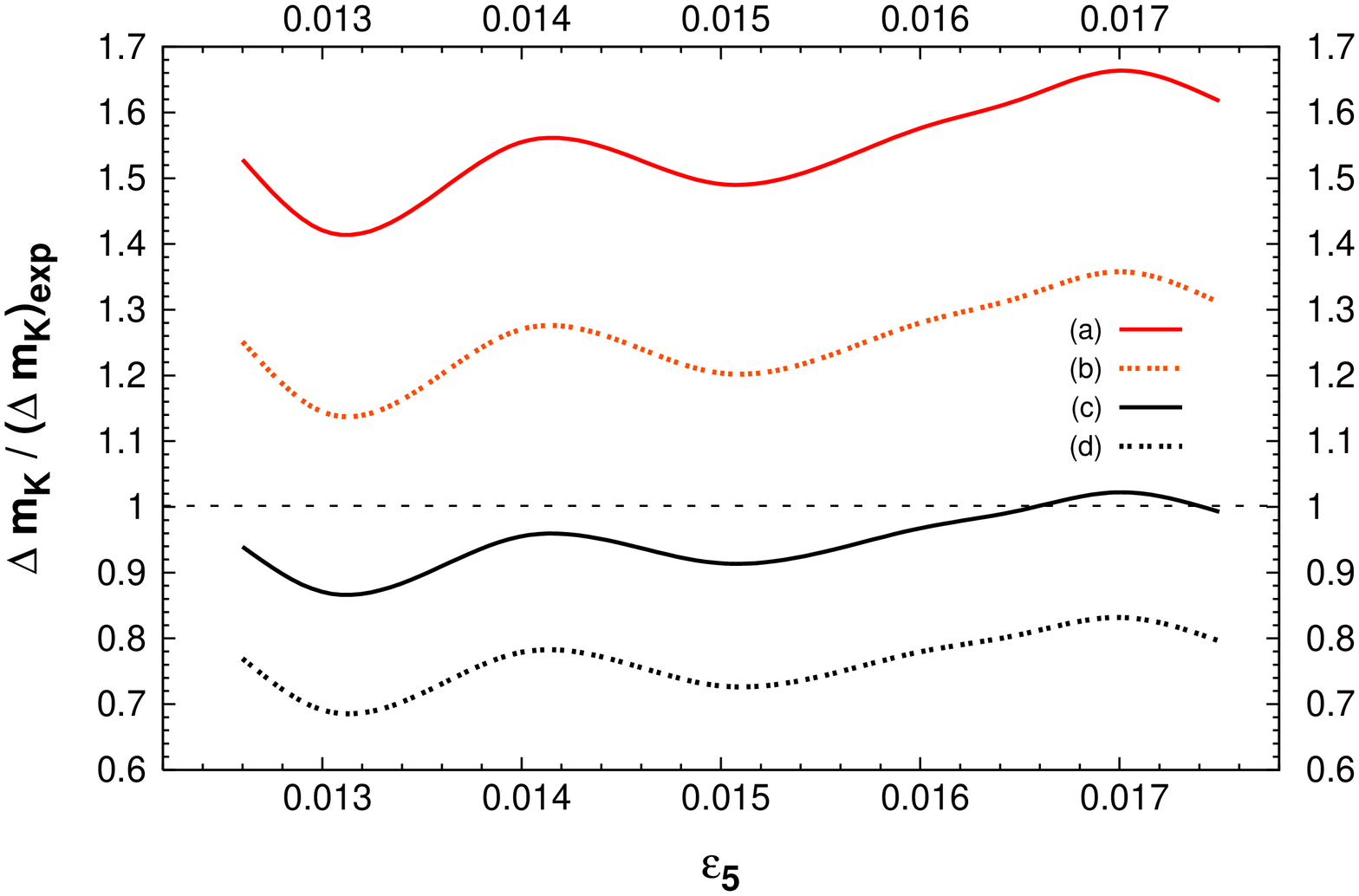,width=50mm,angle=270,clip=} &
\psfig{file=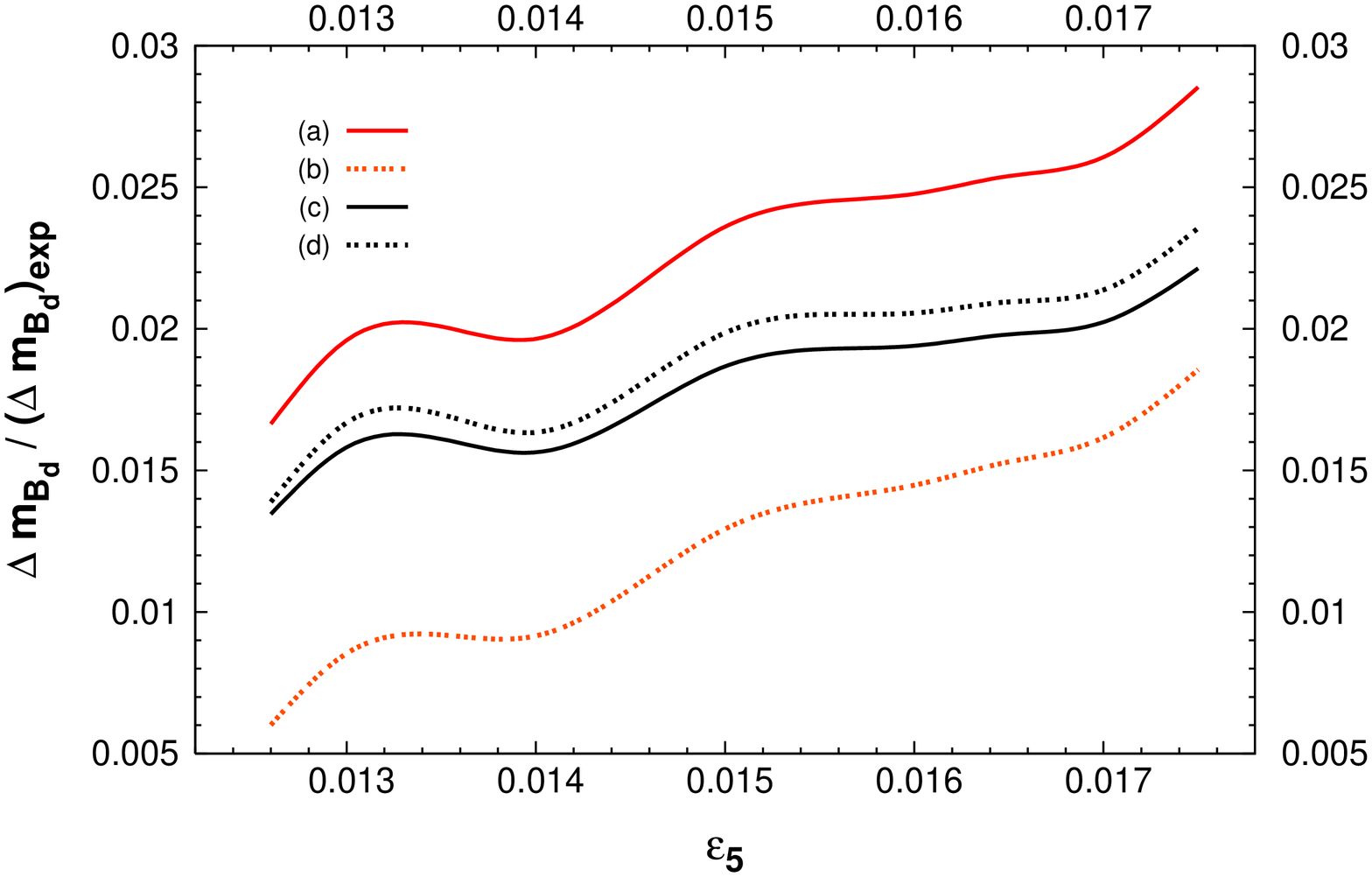,width=50mm,angle=270,clip=} 
    \end{tabular}
    \caption{$\Delta m_K / (\Delta
m_K)_{\text{exp}}$ and $\Delta m_{B_d} / (\Delta m_{B_d})_{\text{exp}}$
as a function of $\varepsilon_5$ for Set B and Higgs textures (a)-(d). } 
    \label{mesonmass}
  \end{center}\vspace*{-4mm}
\end{figure}

A generic analysis of this class of
models (which comprises 21 physical Higgs
states), including minimisation of the scalar potential and derivation
of the mass matrices has been conducted in \cite{Escudero:2005:1}. The
FCNCs induced by the tree level exchange of scalar and pseudoscalar
Higgses has been evaluated, and for example, the contribution to the
neutral scalar to the off diagonal element of the Kaon mass
difference, $ \mathcal{M}^K_{12}$ was found to be
{\small
\begin{equation}\label{MK12:sigma}
\left. \mathcal{M}^K_{12}\right|^\sigma =
\frac{1}{8} \operatornamewithlimits{\sum}_{\begin{smallmatrix}
{j=1-6}
\end{smallmatrix}} \frac{1}{(m^s_j)^2}
\left\{ 
 \left[\operatornamewithlimits{\sum}_{i=1,3,5} 
 \left(
 {\mathcal{V}_d}^{ij*}_{12} \ \ 
 \begin{array}{c} \raisebox{-1mm}{+} \\  \raisebox{+1mm}{-} \end{array}
 \ \ {\mathcal{V}_d}^{ij}_{21} 
\right)
 \right]^2 \langle \overline K^0
 | \ 
 (\bar s \ \ 
\begin{array}{c} 
\raisebox{-.2mm}{1} \\  
\raisebox{+.2mm}{$\gamma_5$} 
\end{array}
 \ \ d) \ (\bar s \ \ 
 \begin{array}{c} \raisebox{-.2mm}{1} \\  
\raisebox{+.2mm}{$\gamma_5$} 
\end{array}
 \ \ d) \
 | K^0 \rangle 
 \right\} \,,
\end{equation}}
where the ``+'' (``-'') sign is associated to the scalar
(pseudoscalar) matrix element, and  
the flavour violation is encoded in
${\mathcal{V}_d}^{ij}_{ab} \propto (V_R^d  Y^{d}_i V^{d \dagger}_L)_{ab}$. 
Assuming four disting textures for the Higgs parameters, that
translate in spectra whose heaviest states are (a): 760 GeV, (b): 880
GeV, (c): 1060 GeV and (d): 800 GeV, we present in Fig.\ref{mesonmass} 
the models contributions to the ratios $\Delta m_K / (\Delta
m_K)_{\text{exp}}$ and $\Delta m_{B_d} / (\Delta m_{B_d})_{\text{exp}}$ 
for $\tan \beta=5$.
Further analysis of other neutral meson systems, as well as that of CP
violation have been also conducted in \cite{Escudero:2005:2}, 
finding that it is possible to have orbifold configurations in
agreement with experimental data for Higgs masses not heavier than the
few TeVs.

\section{Conclusions}
Orbifold models with three Higgs families are very appealing scenarios, 
since the observed structure of quark masses and mixings arises in a
purely geometrical form from superstring constructions. We have
verified that one can easily find orbifold configurations that succeed
in accommodating the experimental data on quark masses and mixings, as
well as neutron meson mass difference and CP violation without the
need to call upon very Higgs textures - $m_h^{\mathrm{max}} \approx 
\mathcal{O}$(few TeV) - to suppress dangerous tree level FCNCs.

\end{document}